\documentstyle[12pt]{article}
\newlength{\extraspace}
\newlength{\extraspaces}
\newcommand{\be}{\begin{equation}
\addtolength{\abovedisplayskip}{\extraspaces}
\addtolength{\belowdisplayskip}{\extraspaces}
\addtolength{\abovedisplayshortskip}{\extraspace}
\addtolength{\belowdisplayshortskip}{\extraspace}}
\newcommand{\ee}{\end{equation}}

\newcommand{\ba}{\begin{eqnarray}
\addtolength{\abovedisplayskip}{\extraspaces}
\addtolength{\belowdisplayskip}{\extraspaces}
\addtolength{\abovedisplayshortskip}{\extraspace}
\addtolength{\belowdisplayshortskip}{\extraspace}}
\newcommand{\ea}{\end{eqnarray}}

\newcommand{\newsection}[1]{
\vspace{3mm} \pagebreak[3] \addtocounter{section}{1}
\setcounter{subsection}{0} \setcounter{footnote}{0}
\begin{center}
{\large {\bf \thesection. #1}}
\end{center}
\nopagebreak
\medskip
\nopagebreak \hspace{3mm}}

\begin{document}
%%%%%%%%%%%%%%%%%%%%%%%%%%%%%%%%%%%%%%%%%%%%%%%%%%%%%%%%%%%%%%%%%%%%%%%%%%%%%
\begin{center}
{{\bf Self Gravitating Instability   of a Compressible  Fluid
Cylinder }}
\end{center}
\centerline{ Ahmed E. Radwan and Gamal G. Nashed}
\bigskip
\centerline{ {\it Mathematics Department, Faculty of Science, Ain
Shams University, Cairo, Egypt }}

\hspace{2cm}
\\
\\
\\
\\
\\
Abstract

The self-gravitating instability of a compressible-inviscid  fluid
cylinder immersed into a self-gravitating tenuous medium of
negligible motion is developed. The stability criterion is derived
based on the linear perturbation technique. Some previous reported
works are recovered. The effect of different factors on the fluid
cylinder instability is discussed. The compressibility has a
tendency for a stabilizing  the model in particular as the sound
speed $a$ is very large in the stable domains $1.0233928\leq
x<\infty$ but comparatively small in the unstable domains
$0<x<1.0233928$ where $x=k R_{_{0}}$ is the dimensionless
longitudinal wavenumber with $k$ is the axial wavenumber and
$R_{_{0}}$ is the radius of the cylinder.  In the absence of the
 compressibility factor the unstable domain is found to be
 $0<x<1.0678$.
\vspace*{1cm}\\
\hspace*{1cm} Key-words: Compressibility, Selfgravitational ,
Instability.
\newpage
\newsection{\bf Introduction}
\vspace*{.5cm}\\
 The instability of a self-gravitating
incompressible fluid cylinder has been investigated for first time
by Chandrasekher and Fermi \cite{CF}. They have utilized the
method of presenting the solinoidal  vectors in terms of poloidal
and toriodal quantities which is valid only for axisymmetric
perturbations. Chandrasekher \cite{C} derived the stability
criterion of such model for axisymmetric and non axisymmetric
modes by using the normal mode analysis technique. Several
extensions for such studies have been carried out upon considering
the effect of different factors on the self-gravitating force
\cite{R, R1}.

The purpose of the present work is to discuss the stability of a
self-gravitating compressible fluid cylinder by utilizing the
normal mode analysis. However, such technique is different from
that used previously by Chandrasekher \cite{C} and Radwan
\cite{R}, because the velocity field in a compressible fluid is no
longer solinoidal, i.e, $\bigtriangledown \cdot {\underline u}
\neq 0$.
\vspace*{1.5cm}\\
\newsection{Basic equations}
 We consider a selfgravitating fluid cylinder of radius
$R_{_{_{_{0}}}}$ embedded into a self gravitating tenuous medium
of negligible motion. The fluid is assumed to be non-viscous and
compressible. The model is acted upon by the self-gravitating and
 pressure gradient forces. We shall use the cylindrical
 coordinates $(r,\varphi,z)$ with the z-axis coinciding with the axis of the
cylinder.

The basic equations appropriate for the present problem are the
equations of motion, continuity equation, equation of state and
the equations satisfying the self-gravitating potentials interior
and exterior the fluid cylinder. These equations could be written
in the form \be \rho \left({\partial \over \partial
t}+({\underline u}\cdot \bigtriangledown) \right){\underline
u}=-\bigtriangledown p+\rho \bigtriangledown V, \ee \be
\left({\partial \over \partial t}+{\underline u}\cdot
\bigtriangledown \right)\rho =- \rho (\bigtriangledown \cdot
{\underline u}), \ee \be \left({\partial \over \partial
t}+{\underline u}\cdot \bigtriangledown \right)p ={\gamma p \over
\rho} \left({\partial \over \partial t} +{\underline u}\cdot
\bigtriangledown \right)\rho, \ee \be \bigtriangledown ^2 V=-4 \pi
G \rho, \ee \be \bigtriangledown ^2 V^e=0, \ee where $\rho,
{\underline u}, V$ and $p$ are the fluid mass density, velocity
vector, self-gravitating potential and kinetic pressure; $V^e$ is
the
 self-gravitating potential of the tenuous medium
exterior the fluid cylinder, $\gamma$ is the ratio of the specific
 heats and G is the gravitational constant. Here the superscript e
 denotes the exterior of the fluid cylinder.
 In the equilibrium state we have
 \begin{center}
  \[{\underline u_{_{_{_{0}}}}}=0, \quad \rho={\rho}_{_{_{_{_{_{_{0}}}}}}},
   \quad  p=p_{_{_{_{_{_{0}}}}}}, \quad V=V_{_{_{_{0}}}},  \quad V^e={V_{_{_{_{0}}}}}^e,\]
\end{center}
where index $0$ characterizes the equilibrium quantities, later on
the index 1 is
 pertaining the perturbed quantities. equations (4)
and (5) in this state read \be \bigtriangledown ^2
V_{_{_{_{0}}}}=-4 \pi G \rho, \ee \be \bigtriangledown ^2
{V_{_{_{_{0}}}}}^e=0. \ee By integrating equations (6) and (7)
with respect to r and determining the constants of integration.
The latter may be determined upon applying the boundary condition
that the self-gravitating potential and its derivative must be
continuous across the boundary surface at $r=R_{_{_{_{0}}}}$.
Consequently, the non-singular solutions of equations (6) and (7)
are given by

\be V_{_{_{_{0}}}}=-\pi G \rho_{_{_{_{_{_{_{0}}}}}}} r^2, \ee \be
{V_{_{_{_{0}}}}}^e=-2 \pi G \rho_{_{_{_{_{_{_{0}}}}}}}
R_{_{_{_{0}}}}^2ln{r \over R_{_{_{_{0}}}}}+C_{_{_{_{0}}}}, \ee
where $C_{_{_{_{0}}}}$ is an arbitrary constant. In the present
initial state, equation (1) yields \be \bigtriangledown
p_{_{_{_{_{_{_{0}}}}}}}=\rho_{_{_{_{_{_{_{0}}}}}}}
\bigtriangledown V_{_{_{_{0}}}}. \ee By integrating equation (10)
and determining the integration constant (note that
$p_{_{_{_{_{_{_{0}}}}}}}=0$ at $r=R_{_{_{_{0}}}}$). Finally, the
distribution
 of the pressure in the unperturbed state is given by
 \be
 {p}_{_{_{_{_{_{_{0}}}}}}}=\pi G \rho_{_{_{_{_{_{_{0}}}}}}}^2(R_{_{_{_{0}}}}^2-r^2).
 \ee
 It is worthwhile to mention here that $p_{_{_{_{_{_{_{0}}}}}}}$ is not constant in
 contrast to other studies for different models acted upon
 capillary or/and electromagnetic forces, \cite{C}, where it is
 found that $p_{_{_{_{_{_{_{0}}}}}}}$ is constant.

\newpage
\newsection{Perturbation Analysis}
For small departure from the equilibrium state, the fluid physical
quantity Q could be expressed as \be Q(r, \varphi, z, t)=\sum_n
\epsilon^n Q_{_{n}}(r, \varphi, z, t), \qquad n=0,1 \ee where Q
stands for $p, {\underline u}, V, V^e$ and  $\rho$. $\epsilon$ is
the amplitude of the perturbation at all times \be
\epsilon=\epsilon_{_{_{_{_{_{_{0}}}}}}} e^{\sigma t}, \ee where
$\epsilon_{_{_{_{_{_{_{0}}}}}}}$($=\epsilon$ at t=0) is the
initial  amplitude and $\sigma$ is the growth rate. If $\sigma$ is
imaginary say $\sigma=i\omega$, $(i=\sqrt{-1})$, then
$\displaystyle{\omega \over 2 \pi}$ is the oscillation frequency.
Based on the expansion (12) and from the view point of the
linearized theory, the deformation along the fluid cylinder
interface due to perturbation could be written in the form
\[
r = R_{_{_{_{0}}}}+R_{_{_{_{1}}}}, \qquad
|R_{_{_{_{1}}}}|<<R_{_{_{_{0}}}},
\]
 with
\be R_{_{_{_{1}}}}= \epsilon e^{i(kz+m\varphi)}. \ee Here k and m
are, respectively, the longitudinal and azimuthal wavenumbers and
$R_{_{_{_{1}}}}$ is the elevation of the surface wave measured
from the initial position at $r=R_{_{_{_{0}}}}$.

By the use of the expressions (12)-(14) for equations (1)-(5), the
relevant perturbation equations are \be \sigma
\rho_{_{_{_{_{_{_{0}}}}}}} {\underline u_{_{_{_{1}}}}} =
-\bigtriangledown p_{_{_{_{_{_{_{1}}}}}}} +
\rho_{_{_{_{_{_{_{0}}}}}}} \bigtriangledown V_{_{_{_{1}}}}, \ee
\be \sigma \rho_{_{_{_{_{_{_{1}}}}}}} =
-\rho_{_{_{_{_{_{_{0}}}}}}}
 (\bigtriangledown \cdot {\underline u_{_{_{_{1}}}}}),
\ee \be p_{_{_{_{_{_{_{1}}}}}}} = \displaystyle {\gamma
p_{_{_{_{_{_{_{0}}}}}}} \over \rho_{_{_{_{_{_{_{0}}}}}}}}
\rho_{_{_{_{_{_{_{1}}}}}}}, \ee \be \bigtriangledown^2
V_{_{_{_{1}}}} = 0, \ee \be
 \bigtriangledown^2 {V_{_{_{_{1}}}}}^e = 0.
\ee

By an appeal to $(\varphi,z)$-dependence (cf. equation (14)) and
based on the linear perturbation technique concerning stability
theory, every perturbed quantity $Q_1(r,\varphi,z,t)$ could be
expressed as $e^{(\sigma t+i(kz+m \varphi))}$ times an amplitude
function of r. Consequently, the self-gravitating equations (18)
and (19) are solved. Upon applying appropriate boundary conditions
across the cylindrical fluid interface at $r=R_0$, the constants
of integration are determined and the non-singular solution of
equations (18) and (19) are given by \be V_{_{_{_{_{1}}}}}=4 \pi G
\rho_{_{_{_{_{_{_{0}}}}}}} R_{_{_{0}}}K_{_{_{m}}}(x)
 I_{_{_{m}}}(kr) e^{i(kz+m\varphi)},
\ee and \be {V_{_{_{_{1}}}}}^e=4 \pi G
\rho_{_{_{_{_{_{_{0}}}}}}}R_{_{_{_{0}}}}I_{_{_{m}}}(x)
K_{_{_{m}}}(kr) e^{i(kz+m\varphi)}. \ee Here $I_{_{_{m}}}$ and
$K_{_{_{m}}}$ are the modified Bessel functions of the first and
second kind of order m while $x(=kR_{_{_{_{0}}}})$ is the
dimensionless longitudinal wavenumber. Combining equations
(15)-(17) and solving the resulting differential equations by
utilizing similar steps as those which are used in solving
equations (18) and (19). The constants of integration could be
identified upon applying the kinematic boundary condition that the
normal component of the velocity vector must be compatible with
the velocity of the perturbed fluid interface. The non-singular
solutions are given by \be {\underline u_{_{_{_{1}}}}}={1 \over
\sigma \rho_{_{_{_{_{_{_{0}}}}}}}}
 \bigtriangledown \left \{
\left[4 \pi G \rho_{_{_{_{_{_{_{0}}}}}}}^2 R_{_{_{_{_{_{_{0}}}}}}}
K_{_{_{m}}}(x) I_{_{_{m}}}(kr)-CI_{_{_{m}}}(r\sqrt{k^2+{\sigma^2
\over a^2}}) \right]e^{i(kz+m\varphi)} \right \}, \ee and \be
p_1=C \ \ I_{_{_{m}}}(r\sqrt{k^2+{\sigma^2 \over a^2}})
e^{i(kz+m\varphi)}, \ee where C is defined by \be
C=\displaystyle{\rho_{_{_{_{_{_{_{0}}}}}}} \left[4 \pi G
\rho_{_{_{_{_{_{_{0}}}}}}}
 x K_{_{_{m}}}(x) I_{_{_{m}}}'(x)-\sigma^2 \right] \over
\sqrt{k^2+\displaystyle{\sigma^2 \over a^2}} \  \ I_{_{_{m}}}
(R_{_{_{_{_{_{_{0}}}}}}} \sqrt{k^2+\displaystyle{\sigma^2 \over
a^2}})}, \ee and $a(=\displaystyle \sqrt{\gamma
p_{_{_{_{_{_{_{0}}}}}}}  \over \rho_{_{_{_{_{_{_{0}}}}}}}} )$ is
the speed of sound in the fluid.

Finally, we have to apply the boundary condition that the normal
component of the stress tensor must be continuous across the fluid
cylindrical interface at $r=R_{_{_{_{0}}}}$. This condition leads
to \be p_1+R_{_{_{_{_{_{_{0}}}}}}} {\partial p_{_{_{_{0}}}}
 \over \partial r}=0, \qquad at \qquad r=R_{_{_{_{0}}}}.
\ee By substituting from equations (11), (23) and (24) into
condition (25), following relation is obtained \be \displaystyle
{\sigma^2 \over 4 \pi G \rho_{_{_{_{_{_{_{0}}}}}}} }=\displaystyle
{y I_{_{_{m}}}'(y) \over I_{_{_{m}}}(y)} \left[
{xK_{_{_{m}}}(x)I_{_{_{m}}}'(x) I_{_{_{m}}}(y)
 \over y I_{_{_{m}}}'(y) }
-{1 \over 2} \right], \ee where \be y=\sqrt{x^2+{\sigma^2 \over
(a/R_{_{_{_{_{_{_{0}}}}}}})^2}}, \ee is the compressible
dimensionless longitudinal wavenumber. Here we get $y=x$ as
$a\rightarrow \infty$.

Equation (26) is the desired self-gravitating dispersion relation
of an inviscid compressible fluid cylinder. It involves the entity
$(4 \pi G \rho_{_{_{_{_{_{_{0}}}}}}})^{-1/2}$ as
 a unit of time, the two kinds
of the modified Bessel functions of different arguments, the sound
speed in the fluid and the classical and compressible wavenumbers
x and y.

In the limiting case as $a\rightarrow \infty$ while $m\geq0$, the
relation (26) reduces to \be \sigma^2=4 \pi G \rho
{xI_{_{_{m}}}'(x) \over I_{_{_{m}}}} \left[I_{_{_{m}}}(x)
K_{_{_{m}}}(x)
 -{1 \over 2}\right].
\ee This relation coincides with the dispersion relation deduced
by Chandrasekhar \cite{C} for solinoidal  velocity vector.

If we suppose that $a\rightarrow \infty$ and $m=0$, the relation
(26) reduces to \be \sigma^2=4 \pi G \rho {xI_{_{_{1}}}(x) \over
I_{_{_{0}}}(x)} \left[I_{_{_{0}}}(x) K_{_{_{0}}}(x)
 -{1 \over 2}\right].
\ee The relation (29) has been derived for first time by
Chandrasekhar and Fermi \cite{CF}. Indeed they have used a
technique which is totally different from that used here. Such a
technique is based on presenting the solinoidal vectors in terms
of poloidal and toroidal quantities.

As the compressibility factor influence is very small, the effect
of the selfgravitating force on the instability of the fluid
cylinder may be determined upon discussing the relation (27) and
(28).

By the aid of the numerical data of the modified Bessel functions
\cite{A}, it is found for all $m\neq0$ that \be
I_{_{_{m}}}(x)K_{_{_{m}}}(x)<{1 \over 2}. \ee Hence \be \sigma^2<0
\qquad for \quad all \qquad m\neq0. \ee Therefore, the fluid
cylinder is gravitationally stable for all purely non-axisymmetric
perturbations. In the axisymmetric perturbation
 mode $m=0$ it is found that
\be \sigma^2<0, \ee in the domain $0<x<1.0667$ while \be
\sigma^2=0, \ee at the critical point $x=1.0667$ and \be
\sigma^2>0, \ee in the wide domain $1.0667<x<\infty$. This means
that the fluid cylinder is self-gravitational unstable as long as
the \be \lambda>(2 \pi/1.0667), \ee while it is stable iff the
perturbed wavelength $\lambda$ satisfy
\[
\lambda\leq(2 \pi/1.0667),
\]
where the equality corresponds to the marginal stability state.

Now, as we have seen in the foregoing discussion the
self-gravitating incompressible fluid cylinder is unstable only in
the axisymmetric mode $m=0$. Therefore, in discussing the general
relation (26) for determining the influence of the
compressibility, we focus our stability discussions for in the
axisymmetric perturbation mode $m=0$. For $m=0$, the relation (26)
reduces to \be \displaystyle {\sigma^2 \over 4 \pi G
\rho_0}=\displaystyle {y I_{_{_{1}}}'(y) \over I_{_{_{0}}}(y)}
\left[ {xk_{_{_{0}}}(x)I_{_{_{1}}}'(x)
 I_{_{_{0}}}(y) \over y I_{_{_{1}}}'(y) }
-{1 \over 2} \right], \ee where the dimensionless compressible
wavenumber  y is still given by (27). The relation (36) has been
calculated in the computer and the effect of different factors are
identified. See figure (1) and table (1). The analytical results
are verified and it is found that the compressibility
 destabilizing the model for different values of $a$ in particular as the sound
 speed $a$ is
very large. It is worthwhile to mention here also that for all
values of $s$ in the range $0.1 \leq s < 5.0$ (with $s=a/
R_{_{0}}$) we found that the model is unstable in the domain
$0<x<1.0233928$. While it is stable in the neighboring domains
$1.0233928 \leq x<\infty$ where the equality corresponds to the
marginal stable state. However, with increasing $s$ values the
areas under the unstable curves are increasing. Corresponding to
$s=0.1, 0.5, 1.0, 2.0$ and 5.0, the maximum modes of instability,
respectively, are $N_{max}=\displaystyle{\sigma \over \sqrt{4 \pi
G \rho_{_{0}}}}=0.066997, 0.246781, 0.309376,0.331429$ and
$0.338533$ at $x=0.46, 0.56, 0.61, 0.61$ and $0.66$. This means
that the compressibility has a tendency of destabilizing the
model. On contrary as $s=0$ we have the selfgravitating unstable
domain is $0<x<1.0678$. This shows that all the unstable domain is
the presence of  compressibility are less than that in the absence
of the  compressibility. This means that the compressibility has a
tendency of stabilizing the model.

\newpage
\begin{center}
\begin{tabular}{|c|c|c|c|c|c|c|} \hline
S & 0.1 & 0.5 & 1.0 & 2.0 & 5.0 \\ \hline x & N & N & N & N & N\\
\hline 0 & 0 & 0 & 0 & 0& 0 \\ \hline
  0.01 &0.0128914  & 0.0153556 &0.0153635 & 0.0153642
& 0.0153643 \\ \hline 0.06 & 0.0351738 & 0.0716648 &0.0725391
&0.0726327 &0.0726493  \\ \hline 0.11 & 0.0456249 &0.114448  &
0.118185 &0.118645 & 0.118733 \\ \hline 0.16 & 0.0525464 &
0.148441 &0.156949  &0.158146 &0.158388  \\ \hline 0.21 &
0.0575092 &0.175619  & 0.190309 &0.192658 &0.193157  \\ \hline
0.26 &0.0611381  &0.197261  &0.219001  &0.222924 & 0.223795 \\
\hline 0.31 & 0.0637511 &0.214264  &0.243448  &0.249344 & 0.250709
\\ \hline 0.36 & 0.0655312 & 0.227276 &0.263907  &0.272127
&0.27411
\\ \hline 0.41 & 0.0665903 & 0.236759 &0.280533  &0.29136 &
0.294075 \\ \hline 0.46 & 0.0669977 &0.24304  & 0.293406 &0.307031
&0.310581  \\ \hline 0.51 & 0.0667935 &0.246338  &0.302542
&0.319054 &0.323518  \\ \hline 0.56 &0.0659956  & 0.246781
&0.307902 &0.327267 & 0.332693 \\ \hline 0.61 & 0.0646025 &
0.244413 &0.309376 &0.331429 & 0.337827 \\ \hline 0.66 &0.0625925
& 0.239189 & 0.306775 &0.331207 &0.338533 \\ \hline 0.71 &
0.0599201 & 0.230963 &0.299799  &0.32614 &0.334293 \\ \hline 0.76
& 0.0565069 &0.21945  &0.287979  &0.315585 &0.324385 \\ \hline
0.81 & 0.0522224 & 0.204152 &0.270569 &0.298585  &0.307766 \\
\hline 0.86 &  0.0468432& 0.184191 &0.246312 & 0.273613 &0.282789
\\ \hline 0.91 & 0.0399449 & 0.157879 &0.212841 &0.23789  &
0.246503\\ \hline 0.96 & 0.0305331 &0.121236  &0.164637 & 0.185057
&0.192222 \\ \hline 1.01 & 0.0143303 & 0.0571351 &0.0780981
&0.0882374 &0.0918586 \\ \hline $\omega^*$& $\omega^*$&
$\omega^*$&$\omega^*$& $\omega^*$ &$\omega^*$ \\ \hline 1.02339&
0.00014708 &0.000581825& 0.000803709& 0.000909179 &0.00085274  \\
\hline 1.06 & 0.0241698 & 0.0967223 &0.132985 &0.150947 &0.157462
\\ \hline 1.11 & 0.0378873 & 0.152123 &0.210249  &0.239625
&0.250416  \\ \hline 1.16 & 0.0484524 &0.19513  & 0.270937
&0.309896 & 0.324356 \\ \hline 1.21 & 0.0576192 & 0.23268&0.324394
& 0.372178& 0.390057 \\ \hline 1.26 &0.0659678 & 0.267053
&0.373646 &0.429795& 0.450929 \\ \hline $x_c=1.023392870924$&0& &
& & \\ \hline $x_c=1.023392849883$&  &0& &  & \\ \hline
$x_c=1.023392870944$& & &0&  &  \\ \hline $x_c=1.023392870827$& &
&&0  &  \\ \hline $x_c=1.023392631646$& & && &0  \\ \hline
\end{tabular}
\end{center}
\begin{center}
Table 1
\end{center}
Stable and unstable domain for selfgravitating compressible fluid
cylinder with $S=\displaystyle{a \over R_{_{_{0}}}}$ and $\omega^*
=\displaystyle {\omega \over \sqrt{4 \pi G \rho_{_{_{0}}}}}$.
\newpage
\begin{center}
Figure (1)
\end{center}
Stable and unstable domains for selfgravitating compressible
cylinder with\\ $S=\displaystyle{a \over R_{_{_{0}}}}$, $N
=\displaystyle {\sigma  \over \sqrt{4 \pi G \rho_{_{_{0}}}}}$ and
$\omega^* =\displaystyle {\omega \over \sqrt{4 \pi G
\rho_{_{_{0}}}}}$.
\end{document}